\documentclass[12pt]{iopart}
\usepackage{graphicx}  % Include figure files

\begin{document}

\title[Proceedings of ``Nuclear Forces and the Quantum Many-Body Problem'']
       {Self-consistent Green's function calculations of ${}^{16}{\rm O}$
       at small missing energies}

\author{   C.~Barbieri\dag\footnote[3]{Email: \tt barbieri@triumf.ca}\ and
        W.~H.~Dickhoff\ddag  
}

\address{\dag\ TRIUMF, 4004 Wesbrook Mall, Vancouver, 
          British Columbia, Canada V6T 2A3}

\address{\ddag\ Department of Physics, Washington University,
	 St.Louis, Missouri 63130, USA}

\begin{abstract}
 Calculations of the one-hole spectral function of ${}^{16}{\rm O}$
for small missing energies are reviewed.
 The self-consistent Green's function approach is employed together
with the Faddeev equations technique in order to study the coupling of both
particle-particle and particle-hole phonons to the single-particle motion.
The results indicate that the characteristics of hole fragmentation
are related to the low-lying states of ${}^{16}{\rm O}$ and an improvement
of the description of this spectrum, beyond the random phase approximation,
is required to understand the experimental strength distribution.
 A first calculation in this direction that accounts for two-phonon
states is discussed.

\end{abstract}



\maketitle

\section{Introduction}

Recent advances in nuclear theory have generated
accurate predictions for the spectrum of most {\em p}
shell nuclei (see, for instance, Refs.~\cite{GFMC,NoCshell}).
 At the same time, other techniques  are becoming available to describe
larger systems and to account for the effects
of the continuum~\cite{CoupClus,GamowSM}.
For medium and heavy systems, relevant information regarding correlations
has been obtained by studying the nuclear spectral function. This was
mainly done by means of variational calculations~\cite{Sick97} and
the self-consistent Green's function (SCGF)~\cite{BaDi04} approaches.

Once a given nuclear Hamiltonian is chosen, quantities such as
the spectroscopic factors are defined uniquely in term of the exact solutions
of the many-body problem. Thus, their knowledge gives direct information
on the correlations induced by that specific nuclear force.
Experimentally, $(e,e'p)$ reactions have provided results for knock out from
orbits both close to~\cite{Louk93,leus} and far from~\cite{Marcel}
the Fermi energy. Although a consistent calculation
(based on the same Hamiltonian) of the initial and final states
has so far been possible only for specific cases~\cite{WimMarco92}, several
analyses~\cite{Paviabook,Udias,LapLi7,Louk93}
suggest that the experimental cross section can be described by standard
phenomenological realistic interactions. This leads to fragments
at small missing energies that have spectroscopic factors of
about~60-70\%~\cite{Louk93}.
Moreover, recent measurements of the spectral function at high missing
energies and momenta~\cite{DanielaPRL} appear to be consistent with
the tail due to the short-range and tensor correlations (SRC) that
are induced by nuclear forces having a repulsive core.  Whether (and how)
softer NN interactions can describe these measurements is an open
(and interesting) question.

For the case of ${}^{16}{\rm O}$, there still exists a substantial 
disagreement between the quenching of spectroscopic factors extracted
from the experiment~\cite{leus,WimMarco92,Udias} and
theory~\cite{muether2h1pO16,GeurtsO16,FaddRPAO16}.
 The latter results suggest that the reasons for this discrepancy
should be looked for in the effects of long-range correlations (LRC) and in
particular in the couplings of single-particle (sp) motion to
low-energy collective excitations.
In this contribution we report about the work done along this line in
Refs.~\cite{FaddRPAO16,2phERPAO16} in order to tackle the above issues
for ${}^{16}{\rm O}$. Sec.~\ref{sec:2} describes the SCGF and Faddeev
formalism employed to couple sp and collective phonons~\cite{FaddRPA1}.
  The results for the hole spectral function
are discussed in Sec.~\ref{sec:3}. These calculations show that a
proper description of the experimental spectral strength requires
an improvement of the spectrum which goes beyond the random phase
approximation (RPA).
 A first step in this direction that includes the propagation two-phonon
states is reported in Sec.~\ref{sec:4}.

\section{Faddeev approach for the single-particle Green's function}
\label{sec:2}

\begin{figure}[t]
 \begin{center}
 \includegraphics[height=1.6in]{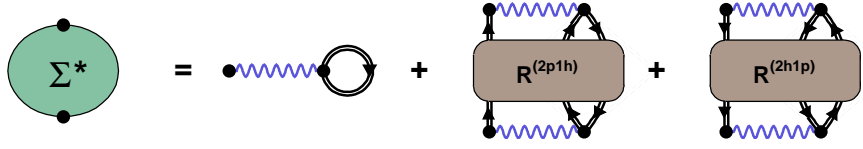}
 \end{center}
\vspace{-0.5in}
\caption[]{
 Diagrams contributing to the irreducible self-energy $\Sigma^*$.
   The double lines represent a dressed propagator and 
   the wavy lines correspond to a G-matrix (that is used in this work as an 
   effective interaction).  The first term is the 
   Brueckner-Hartree-Fock potential while the others represent the 2p1h/2h1p 
   or higher contributions that are approximated through the Faddeev TDA/RPA
   equations.
\label{fig:selfenergy} }
\end{figure}

We consider the calculation of the sp Green's function
\begin{equation}
 g_{\alpha \beta}(\omega) ~=~ 
 \sum_n  \frac{ \left( {\cal X}^{n}_{\alpha} \right)^* \;{\cal X}^{n}_{\beta} }
                       {\omega - \varepsilon^{+}_n + i \eta }  ~+~
 \sum_k \frac{ {\cal Y}^{k}_{\alpha} \; \left( {\cal Y}^{k}_{\beta} \right)^* }
                       {\omega - \varepsilon^{-}_k - i \eta } \; ,
\label{eq:g1}
\end{equation}
from which both the one-hole and one-particle spectral functions, for the
removal and addition of a nucleon, can be extracted.
In Eq.~(\ref{eq:g1}),
${\cal X}^{n}_{\alpha} = {\mbox{$\langle {\Psi^{A+1}_n} \vert $}}
 c^{\dag}_\alpha {\mbox{$\vert {\Psi^A_0} \rangle$}}$%
~(${\cal Y}^{k}_{\alpha} = {\mbox{$\langle {\Psi^{A-1}_k} \vert $}}
 c_\alpha {\mbox{$\vert {\Psi^A_0} \rangle$}}$) are the
spectroscopic amplitudes for the excited states of a system with
$A+1$~($A-1$) particles and the poles $\varepsilon^{+}_n = E^{A+1}_n - E^A_0$%
~($\varepsilon^{-}_k = E^A_0 - E^{A-1}_k$) correspond to the excitation
energies with respect to the $A$-body ground state.
The one-body Green's function can be computed by solving the Dyson equation
\begin{equation}
 g_{\alpha \beta}(\omega) =  g^{0}_{\alpha \beta}(\omega) \; +  \;
   \sum_{\gamma \delta}  g^{0}_{\alpha \gamma}(\omega) 
     \Sigma^*_{\gamma \delta}(\omega)   g_{\delta \beta}(\omega) \; \; ,
\label{eq:Dys}
\end{equation}
where the irreducible self-energy $\Sigma^*_{\gamma \delta}(\omega)$ acts
as an effective, energy-dependent, potential. The latter can be expanded in a
Feynman-Dyson series~\cite{fetwa,AAA} in terms the exact
propagator $g_{\alpha \beta}(\omega)$, which itself is a solution
of Eq.~(\ref{eq:Dys}).
 In this expansion, $\Sigma^*_{\gamma  \delta}(\omega)$ can be represented
as shown in Fig.~\ref{fig:selfenergy} by the sum
of a dressed Hartree-Fock potential and terms that describe the coupling
between the sp motion and more complex excitations~\cite{BaDi04}.
  It is at the level of the 2p1h/2h1p propagator, $R(\omega)$, that
the correlations involving interactions between different collective
modes have to be included.

 The SCGF approach can be initiated by solving the self-energy
and the Dyson Eq.~(\ref{eq:Dys}) in terms of an unperturbed
propagator $g^{0}_{\alpha \beta}(\omega)$. The (dressed) solution 
$g_{\alpha \beta}(\omega)$ is then used to evaluate  an improved 
self-energy, which then contains the effects of fragmentation. The whole
procedure is iterated until self-consistency is reached.
Baym and Kadanoff showed that a self-consistent solution of the
above equation guarantees the fulfillment of the principal
conservation laws~\cite{bk61}.

% The spectroscopic factors $Z_k$ for the removal
%of a nucleon from the $A$ particle system, while leaving
%the residual nucleus in its $k$-th excited state, is obtained from the
% spectroscopic amplitudes ${\cal Y}^k_{\alpha}$. The latter
%are normalized by
%\begin{equation}
% Z_k = \sum_{\alpha}
%\left| {\cal Y}^{k}_{\alpha} \right|^2
%    =  1 + 
%     \sum_{\alpha , \beta}   \left( {\cal Y}^{k}_{\alpha} \right)^* 
%       \left.  \frac{ \partial \Sigma^*_{\alpha \beta}(\omega) }
%                    { \partial \omega}
%       \right|_{\omega = \varepsilon^{-}_k }
%	       {\cal Y}^{k}_{\beta}  \; \; .
%\label{eq:norm}
%\end{equation}
%This result follows directly from the Dyson equation~(\ref{eq:Dys}).
% The same relation applies also to the one-particle spectroscopic
%amplitudes~${\cal X}^{n}_{\alpha}$.

\subsection{Faddeev approach to the self-energy}
\label{sec:2fadd}

 In the following we are interested in describing the coupling of
sp motion to ph and pp(hh) collective excitations of the system.
All the relevant information regarding the latters are included in
the Lehmann representations of the polarization propagator
\begin{eqnarray}
 \Pi_{\alpha \beta , \gamma \delta}(\omega) &=& 
%% g_{\alpha \beta}(\omega) ~=~ 
 \sum_{n \ne 0}  \frac{  {\mbox{$\langle {\Psi^A_0} \vert $}}
            c^{\dag}_\beta c_\alpha {\mbox{$\vert {\Psi^A_n} \rangle$}} \;
             {\mbox{$\langle {\Psi^A_n} \vert $}}
            c^{\dag}_\gamma c_\delta {\mbox{$\vert {\Psi^A_0} \rangle$}} }
            {\omega - \left( E^A_n - E^A_0 \right) + i \eta } 
\nonumber \\
 &-& \sum_{n \ne 0} \frac{  {\mbox{$\langle {\Psi^A_0} \vert $}}
              c^{\dag}_\gamma c_\delta {\mbox{$\vert {\Psi^A_n} \rangle$}} \;
                 {\mbox{$\langle {\Psi^A_n} \vert $}}
             c^{\dag}_\beta c_\alpha {\mbox{$\vert {\Psi^A_0} \rangle$}} }
            {\omega - \left( E^A_0 - E^A_n \right) - i \eta } \; ,
\label{eq:Pi}
\end{eqnarray}
%%\end{equation}
%%%%%%%%%%%%%%%%%%%
%%\begin{equation}
and the two-particle propagator
\begin{eqnarray}
 g^{II}_{\alpha \beta , \gamma \delta}(\omega) &=& 
%% g_{\alpha \beta}(\omega) &=& 
 \sum_n  \frac{  {\mbox{$\langle {\Psi^A_0} \vert $}}
                c_\beta c_\alpha {\mbox{$\vert {\Psi^{A+2}_n} \rangle$}} \;
                 {\mbox{$\langle {\Psi^{A+2}_n} \vert $}}
         c^{\dag}_\gamma c^{\dag}_\delta {\mbox{$\vert {\Psi^A_0} \rangle$}} }
            {\omega - \left( E^{A+2}_n - E^A_0 \right) + i \eta }
\nonumber \\  
&-& \sum_k  \frac{  {\mbox{$\langle {\Psi^A_0} \vert $}}
    c^{\dag}_\gamma c^{\dag}_\delta {\mbox{$\vert {\Psi^{A-2}_k} \rangle$}} \;
                 {\mbox{$\langle {\Psi^{A-2}_k} \vert $}}
                  c_\beta c_\alpha {\mbox{$\vert {\Psi^A_0} \rangle$}} }
            {\omega - \left( E^A_0 - E^{A-2}_k \right) - i \eta } \; .
\label{eq:g2}
\end{eqnarray}
which describe the excited states of the systems with $A$ and
$A \pm 2$ particles, respectively.
%%\end{equation}
\begin{figure}
 \begin{center}
    \parbox[b]{.3\linewidth}{
 \includegraphics[height=2.in]{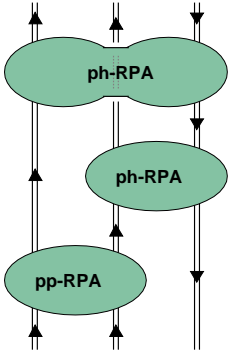} }
    \hfill
    \parbox[b]{.69\linewidth}{
\caption[]{Example of a diagram appearing in the all-orders summation
generated by the set of Faddeev equations.
 \vspace{.4in} 
\label{fig:FaddSum} }  } 
\end{center}
\end{figure}
In general, Eqs.~(\ref{eq:Pi}) and~(\ref{eq:g2})
are the exact solutions of their respective
Bethe-Salpeter equations (BSE).
%, which include correlations beyond
%the Tamm-Dancoff (DTA) and the random phase approximation (RPA) level.
%
 In the calculation of Sec.~\ref{sec:3}, these
have been approximated by solving the dressed Tamm-Dancoff/RPA (DTDA/DRPA)
equations~\cite{schuckbook,DRPAg_papers}, which account for
the effects of the strength distribution of the particle and hole fragments.
The inclusion of correlations beyond RPA is considered in Sec.~\ref{sec:4}.

The ph~(\ref{eq:Pi}) and pp(hh)~(\ref{eq:g2}) propagators are inserted
in the nuclear self-energy by solving a set of Faddeev equations~\cite{gloeck}
for the 2p1h and 2h1p propagators of Fig.~\ref{fig:selfenergy}.
 The details of this approach are given in Ref.~\cite{FaddRPA1}.
 For the present discussion
it is sufficient to note that the motion of three-quasiparticle excitations
is approached in the same way it is normally done for the three-body problem.
 Collective excitations are coupled to sp propagators
generating an infinite series of diagrams, including the one shown
in Fig.~\ref{fig:FaddSum}. This allows to account completely for
Pauli correlations at the 2p1h/2h1p level.

\section{Results for the single-particle spectral function 
          of ${}^{16}{\rm O}$}
\label{sec:3}

 In the calculations described below, the Dyson equation was solved in
a model space consisting of harmonic oscillator sp states.
An oscillator parameter $b =$ 1.76~fm was chosen (corresponding to
$\hbar\omega =$ 13.4~MeV) and all the first four major
shells (from $1s$ to $2p1f$) plus the $1g_{9/2}$ where included.
The results of Refs.~\cite{DRPAg_papers,2phERPAO16}, suggest that
this model space is large enough to properly account for the low-energy
collective states if fragmentaion is accounted for.
 Inside the model space, a Brueckner G-matrix~\cite{CALGM} derived from
the Bonn-C potential~\cite{bonnc} was used as an effective interaction.
The short-range core of this NN interaction induce an additional 10~\%
reduction in the spectroscopic factors~\cite{BaDi04}, which is
accounted for in the solution of the
Dyson equation through the energy dependence of the G-matrix.

\begin{table}[t]
 \begin{center}
 \begin{tabular}{ccccccccccccc}
  \hline 
  \hline 
    Shell      & ~~ & TDA    &~~ & RPA    & ~~& 1st itr. &~ ~& 2nd itr. &~~ & 3rd
    itr. &~ & 4th itr. \\
  \hline 
 $Z_{p_{1/2}}$ & & 0.775  & & 0.745  & & 0.775    & &  0.777   & & 0.774    & & 0.776 \\ \\
 $Z_{p_{3/2}}$ & & 0.766  & & 0.725  & & 0.725    & &  0.727   & & 0.722    & & 0.724    \\
               & &        & &        & & 0.015    & &  0.027   & & 0.026    & & 0.026 \\
%  \hline  
% $n_{d_{3/2}}$ & & 0.038  & & 0.051  & & 0.025    & &  0.025   & & 0.026    & & 0.025    \\
% $n_{d_{5/2}}$ & & 0.035  & & 0.049  & & 0.020    & &  0.021   & & 0.021    & & 0.020    \\
%
% $n_{p_{1/2}}$ & & 0.842  & & 0.821  & & 0.850    & &  0.848   & & 0.848    & & 0.848   \\ 
%
% $n_{p_{3/2}}$ & & 0.872  & & 0.846  & & 0.870    & &  0.871   & & 0.870    & & 0.871   \\ 
%
% $n_{s_{1/2}}$ & & 0.911  & & 0.927  & & 0.925    & &  0.914   & & 0.916    & & 0.930   \\
%  \hline 
%  Total occ.   & & 14.56  & & 14.56  & & 14.56    & &  14.57   & & 14.58    & & 14.63   \\
  \hline 
  \hline 
 \end{tabular}
  \end{center}
    \parbox[t]{1.\linewidth}{
      \caption[]{\label{tab:SpectFact}
  Hole spectroscopic factors ($Z_{\alpha}$) for knockout of a $\ell =1$
      proton from ${}^{16}{\rm O}$.
       The columns `TDA' and `RPA' refer to the initial undressed
      calculations, while the remaining colums resulted from the first
      four iterations of the DRPA equations.
      %% All the values are given as a fraction of the corresponding IPM value.
       Note that the iterated resulted were obtained by constraining
      the lowest $0^+$ solution for ${}^{16}{\rm O}$ at its experimantal
      value, which is at the origin of the fragmentation of the $p_{3/2}$ peak.
%       Also included is the total number of nucleons, as deduced from the
%      complete one-hole spectral function, for each iteration.
      }
      }
\end{table}

\subsection{Effects of RPA correlations and fragmentation}

 For an unperturbed initial propagator the TDA calculation is equivalent
to the one of  Ref.~\cite{GeurtsO16} and yields spectroscopic 
factors equal to 0.775 and 0.766 for the main $p_{1/2}$
and $p_{3/2}$ qhasihole peaks, respectively.
These results are reported in Table~\ref{tab:SpectFact}.
 The introduction of RPA correlations reduces these values and brings them
down to 0.745 and 0.725, respectively.
This shows that collectivity beyond the TDA level is relevant to
explain the quenching of spectroscopic factors.
 We note that due to center-of-mass effects, the above quantities might
need to be increased by about 7\% before they are compared with
the experiment~\cite{O16com}.

\begin{figure}
 \begin{center}
 \includegraphics[height=3.1in]{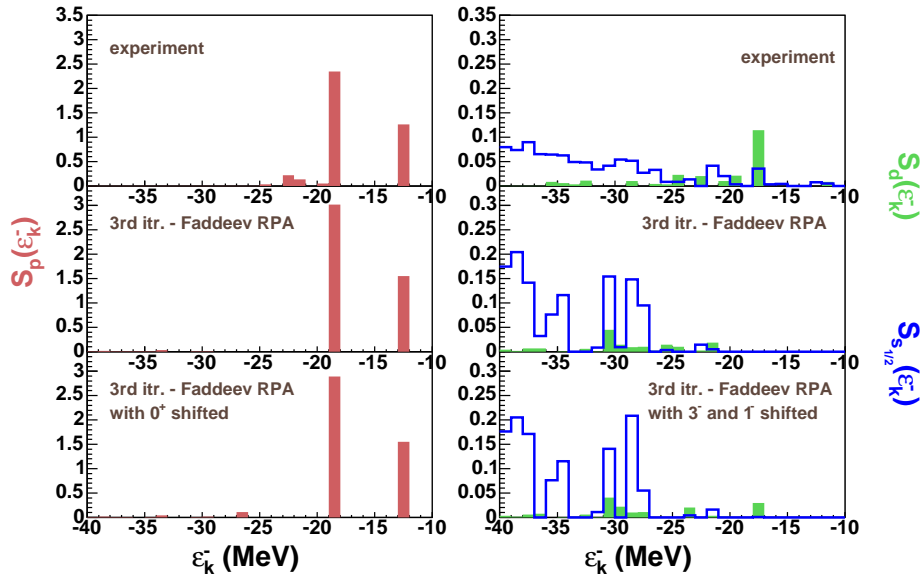}
 \end{center}
%\vspace{-0.5in}
\caption[]{One-proton removal strength as a function of the 
   hole sp energy $\varepsilon^{-}_k = E^A_0 - E^{A-1}_k$
    for ${}^{16}{\rm O}$ for angular momentum
  $\ell =1$ (left) and $\ell =0,2$ (right).
   For the positive parity
   states, the solid bars correspond
   to results for $d_{5/2}$ and $d_{3/2}$ orbitals, while the
   thick lines refer to $s_{1/2}$.  
   The top panels show the experimental values taken from~\cite{leus}.
   The mid panels give the theoretical results for the self-consitent
   spectral function.
   The bottom panels show the results obtained by repeating the 3$^{\rm rd}$
   iteration with a modified ph-DRPA spectrum, in which
   the lowest eigenstates have been shifted to the corresponding
   experimental values.
\label{fig:p+sd_hsf} }
\end{figure}

The RPA results were then iterated a few times to study
the effects of fragmentation.
Since only the low-energy excitations are of interest here,
it is sufficient to keep track only of the largets
fragments that appear ---close to the Fermi energy--- in the
(dressed) sp propagator, Eq.~(\ref{eq:g1}),
while the residual strength is collected in a effective
pole~\cite{jyuan,FaddRPAO16}.
Only a few iterations were required to reach convergence. 
The effect of including fragmentation in the construction of the RPA 
phonons is to increase the strength of the main hole peaks.
  The $p_{1/2}$  strength increases from the 0.745, obtained with the 
undressed input, to 0.776. Analogously the total strength
in the $p_{3/2}$ peak rises to 0.750.
 This behavior is due to the competing effect of the redistribution of the 
strength, which tends to screen the nuclear interaction.
%Nevertheless it is clear  that fragmentation is a relevant feature of nuclear
%systems and that it has to be properly taken into account.

% Together with the main fragments, the Dyson equation produces also a
%large number of solutions with small spectroscopic factors.
%%
% This strength extends down to about -130~MeV and represents 
%the strength that is removed from the main peaks
%and shifted up to medium missing energies.
%
The converged distribution of one-hole strength is shown in the mid panels of
Fig.~\ref{fig:p+sd_hsf}, where it is compared to the experiment (top panels).
The latter is characterized by additional small fragments close to the Fermi
energy, some of which will be discussed in Sec.~\ref{sec:hsf_vs_O16}.
%
%
% Another 10\% of strength is moved to very high energies
%due to SRC~\cite{bv91}. Its location cannot be explicitly
%calculated in the present approach but the effects on the
%reduction of spectroscopic factors at low energy is accounted
%through the
%energy dependence of the G-matrix.
%Accordingly the total number of 
%nucleons deduced from the fully self-consistent spectral function is
%about a 10\% lower than $A =$~16, as reported in Table~\ref{tab:SpectFact}.
% 
 We note that similar results are obtained for the particle strength,
including large peaks near the Fermi level and a fragmented distribution 
at larger energies. This self-energy at positive energies
has been employed recently in studying low energy proton-nucleus
scattering~\cite{BaJenn04}.

\subsection{Role of the lowest excited states in ${}^{16}{\rm O}$}
\label{sec:hsf_vs_O16}

 A deeper insight into the mechanisms that generate the fragmentation
pattern can be gained by investigating directly the connection between
the spectral function and some specific collective states.
 To clarify this point we repeated the above calculations of the
sp propagator by shifting, at each iteration, the solution for the
lowest $0^+$ excitation in ${}^{16}{\rm O}$ to its experimental energy.
 The difference with respect to the preceding  results
is the appearance of a second smaller $p_{3/2}$ fragment at -26.3~MeV,
 which might be interpreted as one of the fragments seen experimentally
at slightly higher energy. This solution arises in the first two iterations
and converges to a spectroscopic factor of 2.6\%,
as seen in Table~\ref{tab:SpectFact}.
 The associated $p$ hole spectral function is shown in the lower-left panel
of Fig.~\ref{fig:p+sd_hsf}.
 This result can be interpreted by considering the $p_{3/2}$ fragments as
generated by holes in the ground state and an excited $0^+$ level of the
${}^{16}{\rm O}$ core.
 If the two levels are close enough in energy, 
the two configurations mix together with the result of fragmenting
the strength over more than one peak.
%We note, however, that the first $0^+$ DRPA solution is of ph nature
%and should does not represent the experimental first excited state.

The other two low-lying states of ${}^{16}{\rm O}$ that may be of some
relevance are the isoscalar $1^-$ and $3^-$, which are reproduced
by RPA type calculations at
$\sim$3~MeV above the experiment (see Fig.~\ref{fig:2PiERPA_3g}
below).
 The lower-right panel of Fig.~\ref{fig:p+sd_hsf} shows the results for the 
even parity spectral functions that are obtained when both
the $3^-$ and the $1^-$ ph-DRPA solutions are shifted to
match their experimental values. 
In this case, a $d_{5/2}$ hole peak is obtained at a missing energy
of -17.7~MeV, in agreement with 
experiment.
% It is interesting to note that the shifting of the $3^-$ and $1^-$
%collective states
%do not produce any other noticeable change in the theoretical spectral
%function.

\section{Two-photon contributions to the spectrum of ${}^{16}{\rm O}$}
\label{sec:4}

\begin{figure}
 \begin{center}
    \parbox[b]{.5\linewidth}{
 \includegraphics[width=1\linewidth]{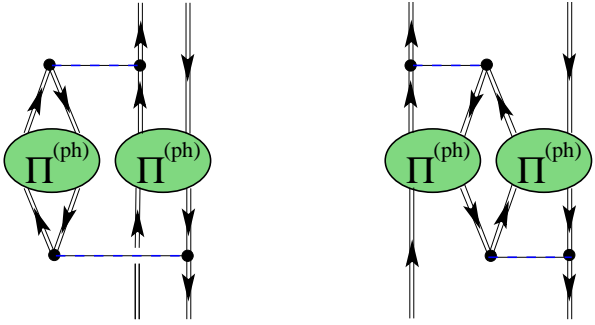} }
    \hfill
    \parbox[b]{.49\linewidth}{
  \caption[]{
 Examples of contributions involving the coupling of two independent
ph phonons. 
 All the diagrams of this type,
obtained by considering all the possible couplings to a ph state,
are included in the BSE kernel by the two-phonon ERPA equations
of Ref.~\cite{2phERPAO16}.
\label{fig:2PiERPA} } }
 \end{center}
\end{figure}

The above results suggest that an improvement of the solution for the spectral
strength would require a better description of the excitation spectrum.
One important problem of (D)RPA
is the appearance of at most one collective phonon for a given $J^{\pi},T$
combination while several low-lying isoscalar $0^+$ and $2^+$
excited states are observed at low energy in ${}^{16}{\rm O}$, as well
as additional $3^-$ and $1^-$ states.
A good description of the spectrum of ${}^{16}{\rm O}$ was obtained
in Ref.~\cite{iach1} by coupling up to four different phonons with
negative parity ($3^-$ and $1^-$).  However, the self-consistent role
of coupling positive parity states and the  dressing of sp propagators
were not investigated. These effects allow for the
partial inclusion of configurations beyond 2p2h
already at the two-phonon level.
Moreover, the inclusion of two-phonon excitations represent the first
correction to the DRPA equation generated by
the Baym-Kadanoff formalism~\cite{wim2ph}.
These conisist of diagrams like the ones of Fig.~\ref{fig:2PiERPA} that
have been included in to the kernel of the BSE.
The relative formalism has been presented in Ref.~\cite{2phERPAO16},
where it is referred to as ``two-phonon extended RPA (ERPA)''.
In this work the ph-DRPA equation has been solved first,
using the self-consistent sp propagator derived in Sec.~\ref{sec:3}.
The lowest DRPA solutions for both the $0^+$, $3^-$ and $1^-$ channels
were shifted down to their relative experimental energies and
then they were employed to generate the two-phonon contributions
for the ERPA calculation.

We note that the solution for the first
isoscalar $0^+$ state in DRPA is found much higher in energy
at $\sim$17~MeV and it has a sharp ph character.
Therefore it cannot be identified with the experimental $0_2^+$ 
state, whose shell model structure is dominated
by 4$\hbar\omega$ configurations~\cite{O16_0+}.
%~\cite{Brown0+,Haxton0+,wbm92}.
%
On the other hand inelastic electron scattering experiments clearly
excite this state~\cite{O16ph86}. 
%%%From the point of view of Green's function theory, the
The one-body response is described by the polarization propagator,
Eq.~(\ref{eq:Pi}), and therefore the total experimental strength must be
represented by $Z_{n_\pi}$, Eq.~(\ref{eq:Z_tot_ph}).
 This indicates a strong coupling to ph configurations (where
``ph'' actually means ``quasiparticle-quasihole'', with bare np-nh
configurations implicitly included by the dressing of the sp propagator). 
 On the basis of this similarity ---and as long as one keeps in mind its
limitations--- it still appears interesting to shift the lowest $0^+$
ph-DRPA solution down in energy and investigate how it mixes with other
configurations.

\begin{figure}[t]
 \begin{center}
    \parbox[b]{.5\linewidth}{
 \includegraphics[width=1.05\linewidth]{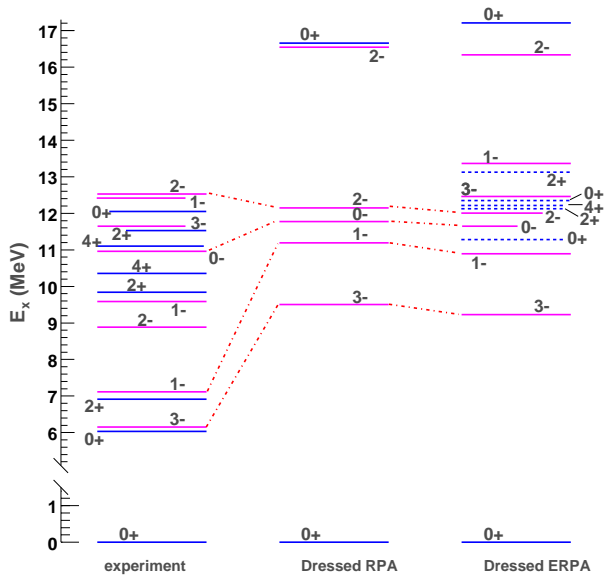} }
    \hfill
    \parbox[b]{.49\linewidth}{
\caption[Two-phonon ERPA results for ${}^{16}\textrm{O}$; dressed input]
 {
 Results for the DRPA and the two-phonon ERPA spectra of ${}^{16}\textrm{O}$
 obtained using the dressed input propagator computed in Sec.~\ref{sec:3},
 middle and right plots, respectively.
  In solving the ERPA equation, the lowest $3^-$, $1^-$ and $0^+$ levels
 of the DRPA propagator where shifted to their experimental energies. All
 other DRPA solutions were left unchanged.
 The ERPA solutions indicated by dashed lines are those with a predominant
 two-phonon chatacter.
  The experimental spectrum is shown on the left. %~\cite{azj}.
\label{fig:2PiERPA_3g} }
 }
 \end{center}
\end{figure}

The ERPA spectrum obtained for  ${}^{16}{\rm O}$ is
displayed in Fig.~\ref{fig:2PiERPA_3g} and Table~\ref{tab:2PiERPA_3g}
together with the total ph strength $Z_{n_\pi}$ of each state,
\begin{equation}
 Z_{n_\pi} ~=~  \sum_{\alpha \beta} ~
      \left| 
       {\mbox{$\langle {\Psi^A_{n_\pi} } \vert $}}
            c^{\dag}_\alpha c_\beta {\mbox{$\vert {\Psi^A_0} \rangle$}} 
      \right|^2   \; \; ,
\label{eq:Z_tot_ph}
\end{equation}
and the relative occupations of ph and two-phonon admixtures in
its wave function.
An isoscalar $0^+$ state with a predominant ph
character is still found at $\sim$17~MeV, as in DRPA, but it is
now characterized by a partial contribution from two-phonon configurations. 
 Table~\ref{tab:2PiERPA_3g} shows that this mixing results in a 
lower solution at $\sim$11~MeV, which is predominately a two-phonon state.
 In both cases the relevant configuration comes from
the coupling of two $0^+$ phonons themselves.
%
% We note that the wave functions for these two
%states contain several relevant ph configurations, obtained
%from different quasiparticle fragments in the $pf$~($sd$) shells combined 
%with quasihole fragments of the $p$~($s$) shells. 
% Therefore, the situation is more complicated than the simple picture
%of only two levels interacting with each other.
%
Of course higher configurations, including three- and four-phonon states,
should be included to reach a complete understanding of the $0^+$ spectrum.
A study of these will be pursued in the future.

\begin{table}[t]
 \begin{center}
 {\small
 \begin{tabular}{cccccccccccccccc}
  \hline 
  \hline 
 $T=0$
   & ~   & \multicolumn{2}{c}{dressed/DRPA} 
   & ~   & \multicolumn{5}{c}{dressed/ERPA}
       &  & $(0_2^+)^2$   & $(3_1^-)^2$
          & $(0_2^+,3_1^-)$ & $(0_2^+,1_1^-)$
          & \\
 $J^\pi$  &  & $\varepsilon^\pi_n$ & $Z_{n_\pi}$ 
          &  & $\varepsilon^\pi_n$ & $Z_{n_\pi}$
          & ~ & $ph$(\%) & $2\Pi$(\%) 
      & ~ & (\%)   & (\%)  & (\%)   & (\%)
          & \\
  \hline 
  \hline 
% $2^+$ &  &  23.77 &   0.468 &   &  23.52 &  0.123 &   & 26   & 74   & & \\
% $2^+$ &  &        &         &   &  22.96 &  0.341 &   & 78   & 22   & & \\
% $2^+$ &  &  20.59 &   0.269 &   &  20.42 &  0.255 &   & 98   &  2   & & \\
%       \\
 $1^-$ &  &        &         &   &  13.37 &  0.148 &   & 21   & 79   & &      &      &      &  79   & \\
 $3^-$ &  &        &         &   &  12.35 &  0.113 &   & 16   & 84   & &      &      & 84    \\
       \\
 $0^+$ &  &        &         &   &  12.15 &  0.001 &   &  1   & 99   & &  3   & 96   \\
 $4^+$ &  &        &         &   &  12.14 &  0.007 &   &  1   & 99   & &      & 99   \\
 $2^+$ &  &        &         &   &  12.12 &  0.008 &   &  1   & 99   & &      & 98   \\
       \\
 $0^+$ &  &  16.62 &   0.717 &   &  17.21 &  0.633 &   & 88   & 12   & &  10  &  0.5 \\
%       \\
 $0^+$ &  &        &         &   &  11.28 &  0.092 &   & 12   & 88   & &  85  &  2   \\
       \\
 $1^-$ &  &  11.19 &   0.720 &   &  10.90 &  0.680 &   & 94.1 &  5.9 & &      &      &      &  5.8  & \\
 $3^-$ &  &   9.50 &   0.762 &   &   9.23 &  0.735 &   & 95.9 &  4.1 & &      &      &  4.0  \\
  \hline 
  \hline 
 \end{tabular}
 }
  \end{center}
    \parbox[t]{1.\linewidth}{
      \caption[Two-phonon ERPA spectrum and strengths; dressed input]
  {
    Excitation energy and total spectral strengths obtained for the 
   principal solutions of DRPA and two-phonon ERPA equations, including
   the total contributions of ph and two-phonon configurations to the ERPA solutions.
    The individual contributions of the relevant two-phonon states
   are also indicated.
    \label{tab:2PiERPA_3g}  }
    }
\end{table}

 The low-lying $3^-$ and $1^-$ states are only slightly
affected by two-phonon contributions and remain substantially
above the experimental energy at 9.23 and 10.90~MeV.
  However, the coupling of these to the $0_2^+$ level reproduce
the second excited states for the same angular momentum and parity.
The two-phonon ERPA approach also generates a triplet of states at 
about 12~MeV with quantum numbers $0^+$,  $2^+$ and $4^+$.
 The solutions for this triplet are almost exclusively
made of $3^- \otimes 3^-$ configurations and therefore have
a 2p2h character.
A similar triplet is found experimentally at 12.05, 11.52
and 11.10~MeV, which correspond to
twice the experimental energy of the first $3^-$ phonon.
%
%The interaction in the pp and hh channels will be needed in order
%to reproduce the correct experimental splitting and strength~\cite{iach1}.

% Of interest are also the $2^+$ solutions that represents the giant quadrupole
%resonance at $\sim$20.7~MeV.
% In this case DRPA and ERPA give 23.7 and 22.9~MeV for the main
%peak but
%with a lower $Z_{n_\pi}$ in the second case. For this state,
%part of the ph strength (about 20\%) is shifted to two-phonon configurations
%representing.
%  This mixing generates the 
%spreading of ph strength over different solutions, consistent with
%the finite width of such a resonant state.
% It should be noted, however, that a discrete basis is used here and therefore
%it is not possible to properly describe the continuous strength
%distribution of a giant resonance, for
%which a continuum or a complex basis should be used~\cite{berggren}.

\section{Conclusions}
\label{sec:concl}

Self-consistent Green's function theory has been applied to study 2h1p
correlations at small missing energies for the nucleus of ${}^{16}{\rm O}$.
The method of the Faddeev equations allows to treat the coupling
of ph and pp(hh) collective modes to the sp motion,
The effects of fragmentation have been included through the dressing
of the sp propagator.
This approach allows to identify the important role played
by the low-lying excited states of ${}^{16}{\rm O}$. These are
essential to generate many of the fragments with small spectroscopic
factors that are seen experimentally, 
examples of which are the $d_{5/2}$ and $p_{3/2}$ states of ${}^{15}{\rm N}$
at 5.20~MeV and $\sim$9~MeV.

The main impediment in obtaining a good theoretical description of
the single particle spectral function of ${}^{16}{\rm O}$ has been identified
in the poor description of the excitation spectrum, as obtained 
by solving the standard (D)RPA equations. We have improved on this by
computing the effects of mixing of ph states with two-phonon configurations.
The results show that these contributions explain the formation
of several excited  states observed at low energy which are not 
obtained by RPA calculations.
 However, it appears that a full solution of the spectrum of ${}^{16}{\rm O}$
with this method requires to consider  up to four-phonon states
and the interaction in the pp and hh channels~\cite{iach1}.

%%%%%%%%%%%%%%%%%%%%%%%%%%%%%%%%%%%%%%%%%%%%%%%%%%%%%%%%%%%%%%%%%%%%%%%%%%
\ack
%One of the authors (C.B.) acknowledges useful discussions
%with J.-M.~Sparenberg and B.~Jennings.
This work was supported in part by the Natural
Sciences and Engineering Research Council of Canada (NSERC) and
in part by the U.S. National Science Foundation
under Grants No.~PHY-9900713 and PHY-0140316.
%%%%%%%%%%%%%%%%%%%%%%%%%%%%%%%%%%%%%%%%%%%%%%%%%%%%%%%%%%%%%%%%%%%%%%%%%%

%%%%%%%%%%%%%%%%%%%%%%%%%%%%%%%%%%%%%%%%%%%%%%%%%%%%%%%%%%%%%%%%%%%%

%%%%%%%%%%%%%%%%%%%%%%%%%%%%%%%%%%%%%%%%%%%%%%%%%%%%%%%%%%%%%%%%%%%%

\end{document}